\documentstyle[11pt]{article}   
\catcode`\@=11
\@addtoreset{equation}{section}
\def\theequation{\arabic{section}.\arabic{equation}}
\def\appendix{\renewcommand{\thesection}{\Alph{section}}\setcounter{section}{0}
              \renewcommand{\theequation}
            {\mbox{\Alph{section}.\arabic{equation}}}\setcounter{equation}{0}}

\def\maketitle{\thispagestyle{empty}\setcounter{page}0\newpage
                \renewcommand{\thefootnote}{\arabic{footnote}}
                  \setcounter{footnote}0}
\renewcommand{\thanks}[1]{\renewcommand{\thefootnote}{\fnsymbol{footnote}}
               \footnote{#1}\renewcommand{\thefootnote}{\arabic{footnote}}}
\newcommand{\preprint}[1]{\hfill{\sl preprint - #1}\par\bigskip\par\rm}
\renewcommand{\title}[1]{\begin{center}\Large\bf #1\end{center}\rm\par\bigskip}
\renewcommand{\author}[1]{\begin{center}\Large #1\end{center}}
\newcommand{\address}[1]{\begin{center}\large #1\end{center}}
\newcommand{\pacs}[1]{\smallskip\noindent{\sl PACS numbers:
                       \hspace{0.3cm}#1}\par\bigskip\rm}
\def\babs{\hrule\par\begin{description}\item{Abstract: }\it} 
\def\eabs{\par\end{description}\hrule\par\medskip\rm}
\renewcommand{\date}[1]{\par\bigskip\par\sl\hfill #1\par\medskip\par\rm}
\newcommand{\ack}[1]{\par\section*{Acknowledgments} #1}

\def\dinfn{Dipartimento di Fisica, Universit\`a di Trento\\ 
                           and Istituto Nazionale di Fisica Nucleare,\\
                                   Gruppo Collegato di Trento, Italia \medskip}

\def\csic{Consejo Superior de Investigaciones Cient\'{\i}ficas (CSIC)\\
 Institut d'Estudis Espacials de Catalunya (IEEC)\\  Edifici Nexus 201, 
                            Gran Capit\`a 2-4, 08034 Barcelona, Spain\\
                               and Departament ECM and IFAE, 
                                 Facultat de F\'{\i}sica,\\
                                      Universitat de Barcelona, 
                                 Diagonal 647, 08028 Barcelona, Spain \medskip}

\newcommand{\guido}{Guido Cognola\thanks{e-mail: \sl cognola@science.unitn.it\rm}}
\newcommand{\sergio}{Sergio Zerbini\thanks{e-mail: \sl zerbini@science.unitn.it\rm}}

\newcommand{\emilio}{Emilio Elizalde\thanks{While on leave at:
Department of Mathematics, Room 2-363A, Massachusetts Institute of Technology,
77 Massachusetts Av., Cambridge, MA 02139-4307, USA;
e-mail: \sl elizalde@ieec.fcr.es\rm\rm}}

\newcommand{\s}[1]{\section{#1}}

\def\hs{\qquad}               
\def\nn{\nonumber}            
\def\beq{\begin{eqnarray}}    
\def\eeq{\end{eqnarray}}      
\def\ap{\left.}               
\def\at{\left(}               
\def\aq{\left[}               
\def\ag{\left\{}              
\def\cp{\right.}              
\def\ct{\right)}              
\def\cq{\right]}              
\def\cg{\right\}}             
\def\R{{\hbox{{\rm I}\kern-.2em\hbox{\rm R}}}}   
\def\H{{\hbox{{\rm I}\kern-.2em\hbox{\rm H}}}}   
\def\N{{\hbox{{\rm I}\kern-.2em\hbox{\rm N}}}}   
\def\C{{\ \hbox{{\rm I}\kern-.6em\hbox{\bf C}}}} 
\def\Z{{\hbox{{\rm Z}\kern-.4em\hbox{\rm Z}}}}   
\def\ii{\infty}                                  
\newcommand{\fr}[2]{\mbox{$\frac{#1}{#2}$}}        
\def\tr{\mathop{\rm tr}\nolimits}                  
\def\Tr{\mathop{\rm Tr}\nolimits}                  
\def\Res{\mathop{\rm Res}\nolimits}                
\def\res{\mathop{\rm res}\nolimits}                
\renewcommand{\Re}{\mathop{\rm Re}\nolimits}       
\renewcommand{\Im}{\mathop{\rm Im}\nolimits}       
\def\dir{/\kern-.7em D\,}                            
\def\al{\alpha}
\def\be{\beta}
\def\ga{\gamma}
\def\de{\delta}
\def\ep{\varepsilon}
\def\ze{\zeta}

\def\la{\lambda}

\def\si{\sigma}

\def\Ga{\Gamma}

\begin{document}

\preprint{UTF }

\title{Dirac Functional Determinants in Terms of the Eta Invariant and the 
Noncommutative Residue}
\author{\guido$^{(1)}$, \emilio$^{(2)}$ 
and \sergio$^{(1)}$ }
\address{${}^{(1)}$\dinfn \\
${}^{(2)}$\csic}

\date{\today}

\babs
The zeta and eta--functions associated with massless and massive Dirac 
operators, in a $D$-dimensional ($D$ odd or even) manifold without boundary, 
are rigorously constructed. 
Several mathematical subtleties involved in this process 
are stressed, as the intrisic ambiguity present in the definition of the
associated fermion functional 
determinant in the massless case and, also, the unavoidable presence 
(in some situations) of 
a multiplicative anomaly, that can be conveniently expressed in terms of the
noncommutative residue. The ambiguity is here 
seen to disappear in the massive case,
giving rise to a phase of the Dirac determinant ---that agrees with very 
recent calculations appeared in the mathematical literature---
and to a multiplicative anomaly 
---also in agreement with other calculations, in the
coinciding situations. After explicit, nontrivial
 resummation of the mass series expansions involving zeta and eta 
functions, the  results are expressed in terms of quite simple formulas.  
\eabs



\s{Introduction}
\label{Form}
It is well known that, within the so-called one-loop approximation in Quantum 
Field Theory, the Euclidean one-loop effective action $\Ga$  may be 
expressed in terms of the
functional determinant of an elliptic differential operator $O$, defined on a 
$D$-dimensional manifold, namely
\beq
\Ga \equiv \ln \det O\,. 
\eeq 
The ultraviolet
one-loop divergences, which  are present, may be regularized by means of the
 zeta-function regularization 
(for recent  reviews, see \cite{eliz94b,eliz95b,byts96-266-1}).
In the bosonic sector, one is dealing, typically, with a second order 
self-adjoint non-negative operator $L$, whose spectrum is bounded from below
(e.g., it is real and non-negative). In this case, a quite elementary 
approach is at our
disposal, namely the  heat-kernel technique, which 
allows to define the zeta-function in an elementary and direct way by means of
the  related Mellin transform, i.e.
\beq
\zeta(s|L)=\frac{1}{\Ga(s)}\int_0^\ii dt \ t^{s-1} \Tr e^{-t L}\,,
\label{mt}
\eeq
valid for $\Re s> D/2$. Here $ \Tr e^{-t L}=\sum_i  e^{-t \la_i}$, $\la_i$ 
being the eigenvalues of $L$. As a 
consequence, $ \ze(s|L)=\sum_i  \la_i^{-s}$.
If zero modes are present, one has to subtract them, replacing 
$\Tr e^{-t L} \to \Tr e^{-t L}-P_0$, $P_0$ being the projector onto the
zero modes. 
The short-$t$ asymptotics of $\Tr e^{-t L}$ have been well studied both in the 
physical and in the mathematical literature \cite{dewi65b,gilk94b}.
For a second-order operator
on a boundaryless $D$-dimensional (smooth) manifold, it reads
\beq
\Tr e^{-tL}\simeq \sum_{j=0}^\ii A_j(L)\
t^{j-D/2}
\:,\label{tas0}
\eeq
in which $A_j(L)$ are the Seeley-DeWitt coefficients, that can be computed
with different techniques.
Making use of the Mellin transform definition, a standard argument leads 
directly to a particular case of Seeley's meromorphic extension theorem,
 namely
\beq
\ze(s|L)
=\frac{1}{\Ga(s)}\aq
\sum_{j=0}^\ii\frac{A_j(L)}{s+j-\frac{D}{2}}+J(s,u)\cq
\:,\label{mero0}
\eeq
where $J(s,u)$ is the analytic part of the zeta-function. It turns out that 
this analytic 
extension is regular at $s=0$ and thus,  the  regularized functional 
determinant
of $L$ can be defined by \cite{ray73-98-154,hawk77-55-133,dowk76-13-3224}
\beq
\log \det L=-\zeta'(0|L)\,.
\label{haw}
\eeq
 
In the fermionic sector, the situation is quite different. Usually, one deals
with a first order  differential operator $A$ (the Dirac operator, the
Rarita-Schwinger operator). To begin with, let us consider the self-adjoint 
case. Here the spectrum is unbounded over the  whole real axis. It is 
obvious that heat-kernel techniques are no longer useful and one has to 
make use of the general theory of complex powers of a 
 pseudo-differential  operator ($\Psi$DO) 
\cite{seel67-10-172}. In the following, we present some  known facts
about the complex powers of  $\Psi$DOs  which we will often use.   
 
Let us consider a $D$-dimensional, smooth (compact) manifold without boundary,
$M_D$, and a (classical) $\Psi$DO, $Q$,
 acting on sections of vector
bundles on $M_D$. To any classical $\Psi$DO, it corresponds a complete
symbol $\si(Q)=Q(x,k)=e^{ikx}Qe^{-ikx}$, such that, modulo infinitely
smoothing operators, one has
\beq
(Q f)(x)\sim \int_{R^D}\frac{dk}{(2\pi)^{D}}\int_{R^D}dy \
e^{i(x-y)k}Q(x,k)f(y)
\:.\label{sy}\eeq
The complete symbol of $Q$ admits an asymptotic expansion for $|k| \to \ii$,
given by
\beq
Q(x,k)\sim \sum_{j=0}^\infty Q_{q-j}(x,k)
\:,\label{sy1}\eeq
where the coefficients fulfill the homogeneity property
$Q_{q-j}(x,tk)= t^{q-j}Q_{q-j}(x,k)$,
for $t>0$, being $Q_{q}(x,k) \neq 0$. The number $q$ is
called the order of $Q$ and $ Q_{q}(x,k)$ its principal symbol. Recall
that $Q$ is elliptic when $ Q_{q}(x,k)$ is invertible.

The complex powers $Q_\theta^{-s}$ can be defined as soon as there exists
a conical neighborough $C_\theta$ of a ray
 $L_\theta=\ag \la \in C, \mbox{arg} \la=\theta \cg$ 
such that it contains, at most, only a finite number of eigenvalues of $Q$. 
Under this condition, the zeta function can be defined, for $\Re s$ larger
than an abscissa of convergence, given by $D/q$, as
\beq
\ze_{\theta}(s|Q)= \Tr  Q_\theta^{-s}=\sum_i \la_i^{-s}\,, 
\label{ze1}
\eeq
where the sum is over the whole spectrum of $Q$ and $\la^{-s}$ is defined 
with respect to the admissible cut $L_\theta$, namely
\beq
\la^{-s}=\exp{(-s \ln_\theta \la)}\,,\hs \theta-2\pi \leq \Im \ln_\theta \la
< \theta\,.
\label{ze11}
\eeq
It is possible to show again that $\ze_\theta(s|Q)$ possesses a meromorphic
continuation to the whole complex plane and that it is regular at the origin
\cite{seel67-10-172}.
Thus, one can define
\beq
\ln \det Q=-\zeta'_\theta(0|Q)\,.
\label{haw11}
\eeq
As a result, in general, both the zeta function  as well as its derivative
may depend on the spectral cut $L_\theta$.   Following Wodzicki, 
one is led to study the function
\beq
\rho(s|Q)=\zeta_\theta(s|Q)-\zeta_{\theta_1}(s|Q)\,,
\label{th}
\eeq
where $L_{\theta_1}$ is another admissible cut. The general properties of 
$\rho(s|Q)$ have been investigated in \cite{wodz82-66-115,wodz84-75-143} and 
they can expressed
by means of the noncommutative residue, that will be introduced in the next 
section. We shall mainly be interested in
$\rho'(0|Q)$, since it is related to the ambiguity in the definition of the
functional determinant of $Q$, and we will discuss it, in an elementary way.
Here we only recall the case when the ambiguity may be present.

Let us introduce the two sectors of the complex plane associated with the two
admissible cuts, namely
\beq
S=\ag \la, \theta <\mbox{arg } \la < \theta_1 \cg\,,\hs
S_1=\ag \la, \theta_1-2\pi  < \mbox{arg } \la < \theta \cg\,. 
\eeq   
Now, it is possible to show that if there is an infinite number of 
eingenvalues of $Q$ in both sectors $S$ and $S_1$, then $\rho'(0|Q)$ is  
non-vanishing and there exists an ambiguity in the definition of the functional
determinant of $Q$ \cite{wodz82-66-115,wodz84-75-143} .

As shown by Wodzicki, a new spectral function plays a role in relation to this
ambiguity. It is the eta-function of $Q$, introduced in Ref. 
\cite{atiy73-5-229}, 
which, for large $\Re s$, is defined by
\beq
\eta(s|Q)=\sum_i \mbox{sign} \la_i |\la_i|^{-s}\,.
\eeq
It can be proven that the eta-function has a meromorphic extension, which is 
regular at the origin \cite{wodz82-66-115,wodz84-75-143} . 

If we apply these general results to the  Dirac operator, it follows
that, in general, there is an ambiguity in the definition of the effective 
action, already noticed  in  \cite{gamboa,dese98-57-7444}, and
 the eta-invariant, i.e. 
$\eta(0|Q)$, will enter the game. This happens only in odd dimension and it 
is known as the appearance of an intrisic parity anomaly term, the so called 
induced Chern-Simons term \cite{niemi,redlich}. 
In even 
dimension the ambiguity is still present, but depends on a Seeley-DeWitt
coefficient and may be absorbed by a redefinition of the scale 
renormalization (see, for example \cite{dese98-57-7444}).        

The contents of the paper are the following. In Sec. 2, some elements of
Wodzicki's theory and the multiplicative anomaly are presented, for the sake 
of completeness. In Sec. 3, 
the zeta-function  and the eta-function associated with 
the massless Dirac operator are revisited. An example is considered in
detail and a brief study of
the relevant Wodzicki function $\rho$ is presented. In Sec. 4 
the massive case is considered and, specifically, the one-loop effective 
action in Sec. 5. The multiplicative anomaly corresponding to the
 very common case of the massive Dirac
operator multiplied by its adjoint is obtained in Sec. 6. 
We also give here explicit formulas for the phase of the massive Dirac 
determinant and for its relation with that of the square root of the 
massive Laplacian. They constitute the main results of the paper.
Some explicit examples are discussed in Sec. 7 in detail.  Sec. 8
is devoted to conclusions. Finally, in an Appendix we analyze the 
disappearance of the ambiguity in the definition of the zeta function 
for the massive Dirac operator, determined by the sign of the mass term. 
This is also, in itself, a remarkable result.

\s{The noncommutative (or Wodzicki) residue}
\label{S:WR}

For the reader's convenience, we will review in this section some 
information concerning the Wodzicki theory on the noncommutative residue
 \cite{wodz87b} (see
also \cite{kass89-177-199,wodz82-66-115,guillem,wodz84-75-143}) 
that will be  used in the rest of the paper.

In order to introduce the definition of the noncommutative residue of 
a $\Psi$DO, $Q$, 
let
us consider an elliptic positive (differential) operator $A$, with $a=2p>q$, 
$p$ integer, and form the 
family of
$\Psi$DOs
$A_Q(u)=A+uQ$, $u$ being a real parameter. The associated zeta-function
reads
\beq
\ze(s|A_Q(u))=\Tr (A+uQ)^{-s}
\:.\label{zq}\eeq
The meromorphic structure of the above zeta-function can be obtained
from the short-$t$ asymptotics of $\Tr e^{-A_Q(u)}$ 
\cite{duis75-29-39},  namely
\beq
\Tr e^{-tA_Q   (u)}\simeq \sum_{j=1}^\ii \al_j(u)
t^{(j-D)/2p}+\sum_{k=1}^\ii \be_k(u)t^k \ln t
\:.\label{tas}\eeq
Note the presence of logarithmic terms  that lead, 
using the Mellin transform, to  double poles
in the meromorphic expansion of $\ze(s|A_Q(u))$, i.e.
\beq
\ze(s|A_Q(u))&=&\frac{1}{\Ga(s)}\at \int_0^1 +\int_1^\ii \ct dt \ t^{s-1}\Tr
e^{-A_Q(u)}\nn \\
&=&\frac{1}{\Ga(s)}\aq
\sum_{j=1}^\ii\frac{\al(u)_j}{s+\frac{j-D}{2p}}-\sum_{k=1}^\ii
\frac{\be_k(u)}{(s+k)^2}+J(s,u)\cq
\:,\label{mero}\eeq
where $J(s,u)$ is the analytic part. Taking the derivative with
respect to $u$ and then the limit $u \to 0$, one gets
\beq
\lim_{u \to 0}\frac{d}{du} \Tr (A+uQ)^{-s}&=&-s\Tr (QA^{-s-1})\nn \\
&=&\frac{1}{\Ga(s)}\aq
\sum_{j=1}^\ii\frac{\al'_j(0)}{s+\frac{j-D}{2p}}-\sum_{k=1}^\ii
\frac{\be'_k(0)}{(s+k)^2}+J'(s,0)\cq
\:.\label{www}\eeq
By definition, the noncommutative residue of $Q$ is given by
\beq
\mbox{res}(Q)=\Res \aq 2p \lim_{u \to 0}\frac{d}{du} \Tr (A+uQ)^{-s}  
\cq_{s=-1}
=2p\be'_1(0)
\:,\label{ncr}\eeq
where $\Res$ is the usual Cauchy residue. It is possible to show
that $\mbox{res}(Q)$ is independent on the elliptic operator $A$ and
that it is a trace in the algebra of classical $\Psi$DOs (actually, the
only trace up to multiplicative constants). From the
above definition and taking the derivative with respect to $u$ at $u=0$
of Eq.~(\ref{tas}),
one obtains a possible way to compute the noncommutative residue.
In fact
\beq
\Tr \at Q e^{-tA} \ct \simeq -\sum_{j \neq D+2p}^\ii \al'_j(0)
t^{(j-D)/2p-1}-\al'_{D+2p}-\frac{\mbox{res}(Q)}{2m} \ln t  +O(t\ln t).
\label{bubu}\eeq
Thus, the noncommutative residue of $Q$ can be read off from the
short $ t$ asymptotics of the quantity $\Tr\at Qe^{-tA}\ct$,
just picking up the coefficient associated with $\ln t$.

When the manifold is non-compact, this is
one of the methods that we have at hand for evaluating the Wodzicki residue,
as long as all the traces involved exist.
For the case of a compact manifold, Wodzicki has obtained a useful
local form of the noncommutative residue, that is,
a density  which can be integrated to yield the noncommutative residue,
namely
\beq
\mbox{res}(Q)=\int_{M_D}\frac{dx}{(2\pi)^{D}}\int_{|k|=1}\tr Q_{-D}(x,k)dk
\:.\label{wod2}\eeq
Here the component of order $-D$ (remember that $D$ is the dimension of
the manifold) of the complete symbol appears as well as the internal trace 
$\tr$.
Eqs. (\ref{www}) or (\ref{bubu}) give \\

\bigskip\noindent{\bf Lemma 1.} In a neighborhood of $z=0$,
\beq 
\Tr (Q A^{-z})= \frac{\mbox{res} (Q)}{2 p z}
+\frac{\ga\:\mbox{res} (Q)}{2p}-\alpha_{D+2p}'(0)+O(z)
\:,\label{A1}\eeq
where $\ga$ is the Euler-Mascheroni constant.
The latter equation gives another way to compute the noncommutative residue,
namely (this expression was also found by Guillemin \cite{guillem})
\beq
\mbox{res} (Q)=2p \aq \Res \Tr (Q A^{-z}) \cq_{z=0}
\:.\label{wodf}\eeq

If $A$ is a positive elliptic (differential) operator of positive order
 $a$, then we
can write
\beq
\ze(A|s+\al)=\Tr[A^{-\al}A^{-s}]=\frac{\res(A^{-\al})}{as}+
\mbox{finite terms}\:,
\hs\hs\al>0\:.
\eeq
Thus,  we see that the residues of the poles of the zeta function
on the positive real axis are proportional to the noncommutative
residues of some negative power of $A$. More precisely we obtain
\beq
\ap\Res\ze(A|s)\right|_{s=\al}=\frac{\res(A^{-\al})}a\:,
\hs\hs\al>0\:.
\label{RES}\eeq
With the latter formula we can compute the residues of all the poles
situated on the positive real axis for the zeta function of any
elliptic operator of positive order.
Eq.~(\ref{RES}) is more general than the well known and well used
heat-kernel counterpart, since it works for operators of even order,
as well as for operators of odd order (for examples and some applications 
see \cite{eliz97-30-2735}).

We conclude this section recalling  the multiplicative
anomaly formula 
(see, for instance, 
\cite{kont93b,eliz98-194-613,eliz98-57-7430,eliz98-532-407}).
We consider two invertible, elliptic, positive (self-adjoint)
operators, $A$ and $B$, on $M_D$ of positive orders, $a$ and $b$,
respectively, and the quantity
\beq
F(A,B)=\frac{\det (AB)}{(\det A)( \det B)}=e^{a(A,B)}
\:.\label{a4}\eeq
It is understood that the functional determinants are here defined by means of 
zeta-function regularization. By construction, $a(A,B)$ is called the 
multiplicative anomaly.
We start from its expression in terms of the Kontsevich-Vishik multiplicative 
anomaly formula \cite{kont93b},
namely
\beq
a(A,B)=\int_0^1 dt \  \mbox{res} \aq \ln \at AB^{-\fr{a}{b}}\ct
 \at \frac{\ln A(t)}{a}-
\frac{\ln [A(t)B]}{a+b}\ct \cq
\:.\label{a10}\eeq
This formula, Eq.~(\ref{a10}), notably simplifies
in the special case of commuting operators. In fact one then has
\beq
a(A,B)=
\frac{b}{2a(a+b)}\mbox{res}\aq (\ln(AB^{-\fr{a}{b}}))^2
\cq
\:,\label{wod3}\eeq
which can be rewritten as the Wodzicki multiplicative formula
\cite{kass89-177-199}
\beq
a(A,B)= \frac{\mbox{res}\aq (\ln(A^bB^{-a}))^2
\cq}{2ab(a+b)}=a(B,A)
\:,\label{wod33}\eeq
where the symmetry property in $A$ and $B$ is manifest.
Physical applications of this formula have been given in 
\cite{eliz98-194-613,eliz98-57-7430,eliz98-532-407}.
A direct computation of the multiplicative anomaly in higher 
dimensional manifolds can be found in \cite{byts98} and an application to 
chiral anomaly has been given in \cite{cogn99-48-375}. 
Very recently, it has been shown that the multiplicative
anomaly gives contributions to the one-loop effective
action obtained by dimensional reduction
\cite{frol99u-86}.

\s{The zeta function of the Dirac operator and its $\eta$ invariant}

As mentioned in the Introduction, if the elliptic operator $A$ is {\it not} 
positive definite, the general theory of complex powers of elliptic 
operators may still be 
used, but then  another spectral function, namely the eta-function plays
(besides the zeta function) a very important role. 
It is convenient to start with the assumption of dealing
with a self-adjoint operator, which we will consider for the moment
to be the massless Euclidean Dirac operator
on a compact (curved) manifold $M$ of dimension $D$.
It has the form
\beq
A=i\dir =i\ga^\mu\nabla_\mu\:,\hs\hs
\ga_\mu=e_k^\mu\ga^k\:,
\eeq
where $e_k^\mu$ are  ``viel-bein'' fields,
$\ga^k$ the Dirac matrices in $D$ dimensions and
$\nabla_\mu$ the covariant derivative in the spin connection.
Since $A^+=A$,  the  
corresponding eigenvalues are  real. Let us denote by $\la_i$ the
positive eigenvalues and by $\mu_i$ the negative ones.  Since the spectrum is
contained in the whole real axis, we have essentially two possible definitions 
of zeta function, because there exist two essentially different cuts, 
one in the lower and the other in the upper half-planes respectively. 
In both cases, the associated sectors 
contain an {\it infinite} number of eigenvalues and the related 
zeta functions read (for $\Re s$ sufficiently large)
\beq
\zeta_+(s|A)=
\sum_i \la_i^{-s}+e^{-i\pi s}\sum_i (-\mu_i)^{-s}
\eeq
and
\beq
\zeta_-(s|A)=
\sum_i \la_i^{-s}+e^{i\pi s}\sum_i (-\mu_i)^{-s}\,.
\eeq
On the other hand, the eta-function may be  also defined as
\beq
\eta(s|A)=
\sum_i \la_i^{-s}-\sum_i (-\mu_i)^{-s}=\Tr \at \frac{A}{|A|}
(A^2)^{-\fr{s}{2}} \ct\,.
\label{eta}
\eeq
Making use of the above definitions, a direct computation gives 
\cite{wodz82-66-115} 
\beq
\eta(s|A)=\frac{1+e^{i\pi s}}{2i\sin \pi s}\zeta_+(s|A)-
\frac{1+e^{-i\pi s}}{2i\sin \pi s}\zeta_-(s|A)
\label{eta1}
\eeq
and
\beq
\zeta(\fr{s}{2}|A^2)=\frac{1-e^{-i\pi s}}{2i\sin \pi s}\zeta_-(s|A)-
\frac{1-e^{i\pi s}}{2i\sin \pi s}\zeta_+(s|A)\,,
\label{zeta1}
\eeq
where $\zeta(z|A^2)$ is the zeta function associated with the spinor
Laplacian $L=A^2$, a second order non negative self-adjoint elliptic operator,
whose zeta function can be unambiguosly defined. 

In order to study the Wodzicki function, namely
\beq
\rho(s|A)=\zeta_+(s|A)-\zeta_-(s|A),
\label{thd}
\eeq
we may express the two zeta functions in terms of the eta-function and
 $\zeta(z|A^2)$. From Eqs. (\ref{eta1}) and  (\ref{zeta1}), one gets
\beq
\zeta_\pm(s|A)=\frac{1}{2}\aq 1+e^{\mp i\pi s} \cq \zeta(\fr{s}{2}|L)+
\frac{1}{2}\aq 1-e^{\mp i\pi s} \cq \eta(s|A)
\label{3.6}\:.
\eeq
As a consequence, we obtain the  result
\beq
\rho(s|A)=i \sin \pi s \aq \eta(s|A)-\zeta(\fr{s}{2}|L)\cq\,.
\label{thd1}
\eeq
For physical applications, it is also useful to give 
the first derivative with respect to $s$, i.e.
\beq
\rho'(s|A)=i \sin \pi s \aq \eta'(s|A)-\frac{1}{2}\zeta'(\fr{s}{2}|L)\cq
+i \pi \cos \pi s \aq \eta(s|A)-\zeta(\fr{s}{2}|L)\cq 
\,.
\label{thd1d }
\eeq
Then, the ambiguity is described by the value at $s=0$ of the Wodzicki 
function and its derivative. These values depend on the regularity at 
$s=0$ of $ \eta(s|A)$ and $\zeta(\fr{s}{2}|L)$. The latter is known to be
regular (the Seeley theorem). With regard to the eta-function, Wodzicki has
proven its analyticity at the origin. However, in the particular but important
case of the Dirac operator, a direct approach is known and we will use it.  

First, let us consider the even dimensional case $D=2p$. Here,
the `$\ga^5$' matrix exists, namely: 
$\ga^{D+1}=\ga^1\cdots\ga^D$, and, 
as is well known, since $\ga^{D+1}$ anticommutes with $\dir$,
 both $\la_i$ and $-\la_i$ are eigenvalues for $\dir$
with eigenfunctions $\psi_i^+$ and $\psi_i^-=\ga^{D+1}\psi_i^+$ respectively.
In these conditions, the eta-function identically vanishes. We have
$\rho(0|A)=0$, and 
\beq
\zeta_{\pm}(s|A)=\frac{1}{2}\aq 1+e^{\mp i\pi s} \cq \zeta(\fr{s}{2}|L)\,,
\label{zetaetapm}
\eeq
\beq
\zeta_{+}(0|A)=\zeta_{-}(0|A)= A_p(L)\,,
\label{mnb}
\eeq
\beq
\zeta'_{\pm}(0|A)= \frac{1}{2}\zeta'(0|L) \mp \frac{i \pi}{2} A_p(L)\,,
\label{zetaetapmp}
\eeq
and
\beq
\zeta'_{+}(0|A)-\zeta'_{-}(0|A)=-i \pi A_p(L)\,.
\label{zetaetapmp11}
\eeq
Here the ambiguity depends on the local functional of the external 
gauge field, the Seeley-DeWitt coefficient  $A_p(L)$, and since the one-loop
effective action is defined modulo counterterms of the same nature, it can 
be reabsorbed by the renormalization procedure.

In the odd dimensional case $D=2p+1$ however, there is no symmetry in the 
spectrum and one has the possibility of 
a non trivial eta-function. Furthermore,
starting from (\ref{eta}), one also may  use the Mellin representation 
of the
eta-function, namely for $\Re s$ sufficiently large
\beq
\eta(s|A)=\frac{1}{\Ga(s)}\int_0^\ii dt\  t^{s-1}
\Tr \at \frac{A}{|A|} e^{-t|A|}\ct\,,
\label{me1}
\eeq
which can be conveniently rewritten as
\beq
\eta(s|A)=\frac{1}{\Ga(\fr{s+1}{2})}\int_0^\ii dt\  t^{\fr{s+1}{2}-1}
\Tr \at A e^{-tA^2} \ct\,.
\label{me}
\eeq
By the standard heat kernel techniques, we can try to find the analytic
continuation of the eta-invariant. To this aim, one needs the
short $t$ asymptotics of the integrand. This involves a highly non trivial 
calculation  and the 
answer is provided by a theorem due to Bismut and Freed 
\cite{bism86-107-103}, which states that
$\Tr \at A e^{-tA^2}\ct =O(t^{1/2})$. More precisely
\beq
\Tr \at A e^{-tA^2} \ct \simeq \sum_{l=0}^\ii C_{l+p+2} t^{l+\fr{1}{2 }}\,,
\label{aa1}
\eeq
where the $C_r$ are suitable coefficients.
Thus, the merophormic extension reads
\beq
\eta(s|A)=\frac{1}{\Ga(\fr{s+1}{2})}\sum_{r=0}^\ii
\frac{C_{l+p+2}}{s/2+l+1} +\eta_0(s|A)\,,
\label{3.17}
\eeq
from which the regularity at $s=0$ of $\eta(s|A)$ directly follows. 
In the latter equation, $\eta_0(s|A)$ is an analytic function of $s$.
As a result,
\beq
\zeta_{+}(0|A)=\zeta_{-}(0|A)=0\,,
\label{amb1}
\eeq
\beq
\zeta'_{\pm}(0|A)= \frac{1}{2}\zeta'(0|L) \pm \frac{i \pi}{2} \eta(0|A)\,,
\label{bbb}
\eeq
and
\beq
\zeta'_{+}(0|A)-\zeta'_{-}(0|A)=i  \pi \eta(0|A)\,.
\label{amb}
\eeq
Here, the ambiguity is non-trivial, since, as it is clear from (\ref{3.17}), 
$\eta(0|A)$ is a non local functional of the external gauge field and it 
cannot be removed by  addition of counterterms. 
Note that, in fact, all our results are particular cases of Wodzicki's 
main theorem and its corollaries and we need no new proofs. The originality 
resides in that we have made them absolutely explicit and applicable to the
very important case of the physics of the Dirac operator. 

Futhermore, from the Bismut and Freed theorem, and making a Mellin transform
inversion, we also  get 
\beq
\Tr \at \frac{A}{|A|} e^{-t|A|}\ct=\frac{1}{2\pi i}\int_{\Re z >0}dz
t^{-z}\Ga(z)\eta(z|A)\,.
\eeq
For $z=0,$ one has a simple pole, while for $z=-2,-4,..$, one has
double poles. Shifting the vertical line to the left, one obtains the 
short $t$ asymptotics
\beq
\Tr \at \frac{A}{|A|} e^{-t|A|}\ct& \simeq& \eta(0|A)\nn\\
&&+
\sum_{l=0}^\ii\frac{2 C_{l+2+p}}{(2l+2)!\Ga(-l-\fr{1}{2})} 
\aq\ga-\Psi(-l-\fr{1}{2})-\ln t\cq t^{2l+2}\,,
\label{boh}
\eeq
where $\Psi(z)$ is the di-gamma function. 
It should be noted the presence of logarithmic terms in this short-$t$
expansion, due to the fact that one is dealing with a $\Psi$DO. From 
Eq.~(\ref{boh}) we have
\beq
\lim_{t \to 0}\Tr \at \frac{A}{|A|} e^{-t|A|}\ct=\eta(0|A)\,,
\label{boh1}
\eeq
which may be useful in the evaluation of the eta invariant.
Another representation of the eta invariant,  which easily follows, is
\beq
\eta(0|A)=\int_0^\ii dt\ 
\Tr \at A e^{-t|A|}\ct\,.
\label{me11}
\eeq
As an application of Eq. (\ref{boh1}), we present a simple proof of the 
classical result obtained in \cite{redlich}, where quantum  massless fermions 
coupled to a classical 
gauge field in $2+1$ dimensions  were shown to  generate a Chern-Simons term,
 confirming the result got in \cite{gamboa85} within a local
zeta-function approach. 
The idea is to compute the first variation of the eta-invariant. Starting from
Eq. (\ref{boh1}), one gets
\beq
\delta \eta(0|A)=-\lim_{t \to 0}\aq t\Tr \at \delta A e^{-t|A|}\ct\cq\,.
\label{boh2}
\eeq
Here $\delta A=\delta (i\ga^\mu\partial_\mu-e\ga^\mu A_\mu)
=-e\ga^\mu\delta A_\mu$, 
$A_\mu$ being the electromagnetic potential. Thus,
\beq
\Tr \at \delta A e^{-t|A|}\ct=-e \int dx  \delta A_\mu \tr
\at \ga^\mu e^{-t|A|}(x,x)\ct\,.
\label{boh22}
\eeq
The spectral theorem gives
\beq
e^{-t|A|}(x,x)=\int_0^\ii d\la e^{-t\sqrt{\la}} \rho_\la(A^2)(x,x)\,.
\eeq
Since one is interested in the  $t \to 0$ limit, only the asymptotics for $\la \to
\ii$ of the local spectral density is relevant. In $D=3$, one has (see, for
example, \cite{cogn89}) when $\la \to \ii$
\beq
\rho_\la(A^2)(x,x)\simeq \frac{\la^{\fr{1}{2}}}{\Ga(\fr{3}{2})}+
a_1(x) \frac{\la^{-\fr{1}{2}}}{\Ga(\fr{1}{2})}+O(\la^{-\fr{3}{2}})\,,
\eeq 
where $a_1(x)=\frac{\ga^\nu F_\nu^*}{(4 \pi)^{3/2}}$, with 
$F_\mu=\ep_{\mu \nu \rho}F^{\nu \rho}$, is the first non-trivial 
Seeley-DeWitt coefficient releted to the spinor Laplacian.
The leading term gives no contribution and taking the internal traces, one has
for the non trivial ground state fermion current contribution
\beq
\frac{\delta \eta(0|A)}{\delta A_\mu}=-\frac{e}{2\pi^2}F_\mu^*\,,
\eeq
which is the Redlich result. We note that the sign ambiguity is always 
present
in the massless case, because the total effective action is given by 
(\ref{bbb}) and has nothing to do with the Pauli-Villars 
regularization, as often claimed in the literature. 

We close this section by presenting  a heuristic argument for the regularity 
of the eta-function at the origin. Let us start from (\ref{eta}), namely
\beq
\eta(s|A)=\Tr \at Q
(A^2)^{-\fr{s}{2}} \ct\,,
\label{eta8}
\eeq
with $Q=\frac{A}{|A|}$, $|A|=\sqrt{A^2}$. 
$Q$ is a  $\Psi$DO of zero  order. By making use of
{\it Lemma 1}, near the origin we have
\beq
\eta(s|A)=\frac{\mbox{res} (Q)}{s}+ \frac{\ga}{2} \mbox{res}( Q)-
\alpha'_{D+2}(0)+O(s)\,.
\label{etap}
\eeq
Thus, we have arrived at the following statement:
\begin{description}
\item{\bf Proposition 1.}
The eta-function related to a non positive elliptic operator $A$  is regular
at $s=0$ if and only if $\mbox{res} (\frac{A}{|A|})=0$, where
$|A|=\sqrt{A^2}$.
\end{description}

{\bf Example 1.}
On a compact flat manifold $M$ with odd dimension $D$, the massless 
Dirac operator has the form
\beq 
A=i\dir =i\ga^\mu\nabla_\mu\:,\
\eeq
$\ga^\mu$ being the Dirac matrices in $D-1$ dimensions. For the sake of 
simplicity, let us work in $D=3$ and
assume a constant gauge potential $V_\mu$, namely a vanishing field 
strenght
$F_{\mu \nu}$. Thus
\beq
A^2=-\dir^2=-\Box\,,\hs\hs
|A|=\sqrt{-\Box} \,.
\eeq
The local symbol reads
\beq
\sigma \at\frac{ A}{|A|}\ct(x,k) =
\frac{ k\!\!\!/+V\!\!\!\!\!/}
{\sqrt{k^2+V^2+2 k\!\!\!/ V\!\!\!\!\!/ }}\:,
\eeq
In order to select the homogeneous component of degree $-3$, we may 
consider $k \to \la k$, with $\la$ very large, and pick up the term 
proportional to $1/\la^3$. One gets, with $|k|=1$,
\beq
\sigma \at \frac{ A}{|A|} \ct(x,\la k)=...+
\at C_1  V\!\!\!\!\!/ +C_2 ( V\!\!\!\!\!/)^3 \ct \la^{-3}+... 
\eeq
where the $C_1$ and $C_2$ are simple functions  of $V$.
In the evaluation of the Wodzicki residue one has to take the 
internal trace $\tr$ and this gives a vanishing contribution because,
in $D=3$, not only $\tr  V\!\!\!\!\!/=0$, but  also  $\tr 
({V\!\!\!\!\!/ \ }^3)=0$, 
since $\tr \ga_\mu \ga_\nu \ga_\rho=2\ep_{\mu \nu \rho}$ . 

\s{The massive Dirac operator}

Let us  consider now  the case of massive spinors.
Denoting the massive operator by $K=A+im$, we have
\beq
K\psi_i=(A+im)\psi_i=(\la_i+im)\psi_i\:,
\eeq 
where, as above, $\la_i$ and $\psi_i$ are the eigenvalues and
eigenfunctions of $A$. $K$ is non-hermitian and its eigenvalues are now 
complex numbers. As in the massless case, we also have here two possible 
spectral cuts, and the two zeta functions may in principle be defined  as
\beq
\ze_{\pm}(s|K)&=&\sum_{\la_i>0}(\la_i+im)^{-s} +\left. \sum_{\mu_i<0}\aq 
e^{\mp i \theta}(-\mu_i)+im \cq^{-s} \right|_{\theta =\pi} \,.
\eeq

However, due to the presence here of the mass term $+im$ (and being the
mass $m$ positive), in order to avoid
a singularity in the complex powers of the spectrum of $K=A+im$ when we
perform the analytical continuation of the spectral values (from positive
$-\mu_i$ to negative $\mu_i$, along a path $\Gamma_\pm$: $-\mu_i e^{\pm i
\theta}, 0\leq \theta \leq \pi$),
it is compulsory to take $\ze_+(s|K)$ ---e.g.,
to perform the analytic continuation in the upper half plane.
In fact, a singular point is lying in the lower half complex $\mu$-plane,
namely at $-im$, that gives rise to a cut in the complex $s$-plane
when performing then the analytic continuation in $s$. Note that we have
to perform two different analytical continuations here, in a compatible 
way, and this is actually guaranteed in our method (see a more detailed 
discussion of this issue in the Appendix). Quite on
the contrary, when we will consider, below, the adjoint operator $K^+=A-im$,
the analytical continuation must be performed on the lower half-plane,
namely the relevant zeta function will be $\ze_-(s|K^+)$,
in order to avoid the singularity (see also the Appendix). 
As we shall prove in a subsequent section,
this argument turns out to be very important for the obtaintion of the correct
value for the multiplicative anomaly corresponding to the product of the
two operators,  $K$ and $K^+$. Notice also that our definition of the zeta
function for the operators  $K$ and $K^+$ slightly differs from the one adopted
in \cite{dese98-57-7444}, where the imaginary part of the eigenvalues was also
rotated.\footnote{To compare,  the definition in \cite{dese98-57-7444}
corresponds to rotating $(-\mu_i-im) e^{\pm i \theta}, 0\leq \theta \leq
\pi$, from $\theta =0$ to $\theta =\pi$.}  We
believe that our definition is the most natural one in the massive case,
being the minimal extension of the corresponding definition for selfadjoint
operators with (in general) positive and negative spectrum ---that was on its
turn obtained from the rigorous mathematical theory for elliptic selfadjoint
operators, with no negative eigenvalues.

If the mass is sufficiently small, binomial expansion gives
\beq
\ze_+(s|K)=\sum_{k=0}^{\ii}
\frac{(-1)^k\Ga(s+k)\:(im)^k}{\Ga(s)k!}
\:\ze_+(s+k|A)\,,
\eeq
and analogously for $\ze_-(s|K^+)$.
Making use of Eqs. (\ref{3.6}), one arrives at
\beq
\ze_+(s|K)&=&\frac12[1+e^{-i\pi s}]\:\ze(\fr{s}{2}|L)+
\frac12[1-e^{-i\pi s}]\:\eta(s|A)\nn \\
&&\hs+
\sum_{k=1}^{\ii}\frac{(-1)^k\Ga(s+k)\:(im)^k}{2\Ga(s)k!}
\:\ag[1+e^{-i\pi(s+k)}]\:\ze(\fr{s+k}{2}|L)\cp\nn\\
&&\hs\hs+\ap[1-e^{-i\pi(s+k)}]\:\eta(s+k|A)\cg\:.
\label{mass}
\eeq
We have thus obtained, in a direct way, the meromorphic extension of
the zeta function associated with the massive Dirac operator,
starting from the knowledge of the massless one. 

As an example, if $M$ is a flat even-dimensional manifold, then $\ze(s|L)$
has only a simple pole at $s=D/2$ with residue equal to
$(4\pi)^{-D/2}D{\cal V}$, ${\cal V}$ being the volume of $M$.
This means that $\ze(s|K)$ has a finite number of
simple poles at the points $s=D-k>0$ ($k=0,1,...,D-1$),
whose residues read
\beq
\ap\Res\ze_+(s|K)\right|_{s=D-k>0}=
\frac{2{\cal V}\Ga(D+1)\:\:(-im)^k}{(4\pi)^{D/2}\Ga(D/2)\Ga(D-k)k!}\:.
\label{resDir}
\eeq
Observe that this result is independent of the choice of definition for the 
zeta function ($\ze_\pm$).

Furthermore, consider the free massive Dirac operator
in a flat manifold.  For the symbol, we have
\beq
\si(i\dir+im)= p\!\!/+im\:,\hs\hs
\si(i\dir+im)^{-\al}=\frac1{ p\!\!/^\al}\:
\aq\sum_{k=0}^\ii\:\at\frac{-im}{ p\!\!/}\ct^k\cq^\al
\eeq
and, in order to compute the non-commutative residue, we must pick
up the coefficient of $ p\!\!/^{-D}$ in the previous expansion.
We see that such a coefficient is non-vanishing
only when $\al=n$ is an integer number smaller than or equal to $D$.
Under such condition, we get
\beq
\res([A+im]^{-n})=
\frac{\Ga(D+1)\:\:(-im)^{D-n}}{(4\pi)^{D/2}\Ga(D/2)\Ga(n)(D-n)!}\:.
\eeq
Using this result in Eq.~(\ref{wod2})
we obtain Eq.~(\ref{resDir}) again.

\s{The one-loop effective action }
\label{S:examples}
 
In this section we will compute the effective action for a massive spinor
field on a $D$-dimensional manifold in terms of the zeta-function and the
eta-function
of the related massless operator. The effective action turns out to be 
proportional to the logarithm of the functional determinant of the Dirac 
operator, which may be expressed as minus the first derivative of the 
zeta function at zero. The
massive Dirac operator $K=A+im$  is non-hermitian, and one has
$K^+=A-im=K-2im$. However, $K^+K=L+m^2$ is a positive self-adjoint
second order Laplace-like operator, and  binomial expansion
gives
\beq
\zeta(s|K^+K)=\ze(s|L+m^2)=\ze(s|L)+\sum_{j=1}^{\ii}(-1)^{j}
\frac{\Ga(s+j)m^{2j}}{\Ga(s)j!}\ze(s+j|L)\,.
\label{999}
\eeq
It is plain that the functional determinant associated with $L+m^2$ will 
play here an important role.

Let us begin with  the even dimensional case, i.e. $D=2p$. Recall that, in 
this situation, 
$\eta(s|A)=0$. Splitting the sum in Eq. (\ref{mass}) into even and odd terms
and making use of Eqs. (\ref{3.6}), it is easy to show that 
the latter gives a contribution of order $O(s^2)$. We can isolate the 
singular term by writing
\beq
\ze_{+}(s|K)=\frac12 [1+e^{- i\pi s}]\:\ze(\fr{s}{2}|L)+
\frac{1}{\Ga(s)}\at\frac{G_+(s)}{s}+\phi_+(s)\ct+O(s^2)\,,
\label{kkk}
\eeq
and
\beq
\ze_{-}(s|K^+)=\frac12 [1+e^{ i\pi s}]\:\ze(\fr{s}{2}|L)+
\frac{1}{\Ga(s)}\at\frac{G_-(s)}{s}+\phi_-(s)\ct+O(s^2)\,,
\label{kkk-}
\eeq
where
\beq
\phi_\pm(s)&=&\frac12\sum_{j=1}^{p}\frac{(-1)^j\Ga(s+2j)\: m^{2j}}{(2j)!}
\:[1+e^{\mp i\pi s}]\:\aq\ze(\fr{s}{2}+j|L)
-\frac{2A_{p-j}(L)}{s\Ga(\fr{s}{2}+j)} \cq\nn\\
&&\hs\hs+
\frac12\sum_{j=p+1}^{\ii}\frac{(-1)^j\Ga(s+2j)\: m^{2j}}{(2j)!}
\:[1+e^{\mp i\pi s}]\:\ze(\fr{s}{2}+j|L)
\label{mass2}
\eeq
and
\beq
G_\pm(s)=\sum_{j=1}^{p}\frac{(-1)^j\Ga(s+2j)\: m^{2j}}{(2j)!}
\: [1+e^{\mp i\pi s}]\:
\frac{A_{p-j}(L)}{\Ga(\fr{s}{2}+j)} \,.
\label{mass23}
\eeq
It is easy to see that both $\phi_\pm(s)$ are regular at $s=0$. 
Taking derivatives at $s=0$, we get
\beq
\ze_{+}'(0|K)=\frac12 \aq \:\ze'(0|L)- i\pi\ze(0|L)\cq
+\ga G_+(0)+G_+'(0)+\phi_+(0)\,,
\label{kkk11}
\eeq  
and a similar expression for $\ze_{-}'(0|K^+)$.

On the other hand, from (\ref{999}), one readily has
\beq
\frac{1}{2}\ze'(0|L+m^2)=\frac12 \:\ze'(0|L) 
+\ga G(0)+\phi(0 ) \,,
\label{kkk9}
\eeq
where $G(0)=G_\pm(0)$ and analogously $\phi(0)=\phi_\pm(0)$.
Comparing Eqs.~(\ref{kkk11}) and (\ref{kkk9}) we finally obtain
\beq
\ze_{+}'(0|K)&=&\frac12\aq\:\ze'(0|L+m^2)- i\pi\ze(0|L+m^2)\cq
+\sum_{j=1}^{p}\frac{(-1)^j c_j m^{2j}A_{p-j}(L)}{j!}\:\,,
\label{kkk99}
\eeq
and
\beq
\ze_{-}'(0|K^+)&=&\frac12\aq\:\ze'(0|L+m^2)+ i\pi\ze(0|L+m^2)\cq
+\sum_{j=1}^{p}\frac{(-1)^j c_j m^{2j}A_{p-j}(L)}{j!}\:\,,
\label{kkk99-}
\eeq
with
\beq
c_j=\sum_{l=1}^j \frac{1}{2l-1}\,.
\eeq
The above formula may be useful in the actual evaluation of the 
derivative of the zeta function of the massive Dirac operator. 
 Here, again, the ambiguity 
depends on the Seeley-DeWitt coefficient, $\ze(0|L+m^2)$, only.
We should point out  that, as it stands, the above equation is valid for 
arbitrary, but finite, values of the mass $m$. This observation allows us to
obtain a large mass expansion, simply by recalling an asymptotic theorem due to
Voros \cite{voros}, which states that, for large $m$, the asymptotics of the
massive Laplacian are determined by the short asymptotics of its heat-kernel
trace. As a result, we have
\beq
\ze(s|L+m^2)\simeq \sum_{r=0}^\ii\frac{A_r(L)\Ga(r-p+s)}{\Ga(s)}
m^{2(p-r-s)}\simeq \sum_{r=0}^p \frac{(-1)^{p-r}}{(p-r)!}A_rm^{2p-2r-2s}+O(s)\,,
\label{mn}
\eeq
and
\beq
\ze'(0|L+m^2)&\simeq& \sum_{r=0}^{p-1}\frac{(-1)^{p-r}A_r(L)}{(p-r)!}m^{2(p-r)}
\left( -\ln m^2+\sum_{j=1}^{p-r}\frac{1}{j} \right)-A_p\ln m^2  \nn \\
&&\hs\hs+\sum_{r=p+1}^\ii A_r(L) \Ga(r-p)m^{2(p-r)}
\,.
\label{nmnm}
\eeq

To summarize, we have reduced the problem of the first order masiive Dirac 
Operator (in the even dimensional case) to the much more 
familiar one for the simple Laplacian operator $L+m^2$, and traced back
the terms coming from the previous ambiguities (fixed for $m\neq 0$) of the 
zeta  function definition,
to the term involving the Seeley-DeWitt coefficient $\ze(0|L+m^2) $. 
Our results are in agreement with the ones obtained in 
\cite{dese98-57-7444},
with a completely different method. 

In the odd dimensional case, namely $D=2p+1$, the eta invariant is  
non-vanishing and we 
have to refer to its meromorphic continuation 
given by the Eq. (\ref{3.17}).
As a consequence, 
\beq
\ze_{+}'(0|K)&=&\frac12\:\ze'(0|L)+\frac{i\pi}{2}\eta(0|A)+
\sum_{j=1}^{\ii}\frac{(im)^{2j}}{2j}\ze(j|L)\nn \\
&&\hs-\sum_{j=0}^{\ii}\frac{( im)^{2j+1}}{2j+1}\eta(2j+1|A)
+\frac{i\pi}{2}\sum_{j=0}^{p}\frac{(im)^{2j+1}}
{\Ga(j+\fr{3}{2})}A_{p-j}(L)\,,
\label{kkk1}
\eeq
and
\beq
\ze_{-}'(0|K^+)&=&\frac12\:\ze'(0|L)-\frac{i\pi}{2}\eta(0|A)+
\sum_{j=1}^{\ii}\frac{(im)^{2j}}{2j}\ze(j|L)\nn \\
&&\hs-\sum_{j=0}^{\ii}\frac{(- im)^{2j+1}}{2j+1}\eta(2j+1|A)
-\frac{i\pi}{2}\sum_{j=0}^{p}\frac{(im)^{2j+1}}
{\Ga(j+\fr{3}{2})}A_{p-j}(L)\,.
\label{kkk1--}
\eeq
If we introduce the eta-like functions
\beq
\eta(s|A\pm im)=
\sum_i (\la_i\pm im)^{-s}-\sum_i (-\mu_i\pm im)^{-s}\,,
\label{etam}
\eeq
for small mass, we have
\beq
\eta(s|A \pm im)=\eta(s|A)+
\sum_{k=1}^{\ii}\frac{(\mp im)^{k}\Ga(s+k)}{\Ga(s)k!}\eta(s+k|A)
\,.
\label{eta+-}
\eeq
Thus, the meromorphic properties of $\eta(s|A \pm im)$ follow from the ones
of the eta function. In particular, one obtains
\beq
\eta'(0|A+im)-\eta'(0|A-im)=-2\sum_{j=0}^{\ii}
\frac{( im)^{2j+1}}{2j+1}\eta(2j+1|A)\,.
\label{bo}
\eeq
As a result, we have
\beq
\ze_{+}'(0|K)&=&\frac12\:\ze'(0|L+m^2)+\frac{i\pi}{2}\eta(0|A)
\nn\\
&&+\frac{1}{2}\aq\eta'(0|A+im)-\eta'(0|A-im)\cq
-\frac{\pi}{2}\sum_{j=0}^{p}\frac{(-1)^j 
m^{2j+1}}{\Ga(j+\fr{3}{2})}A_{p-j}(L)\,,
\label{kkk111}
\eeq
and
\beq
\ze_{-}'(0|K^+)&=&\frac12\:\ze'(0|L+m^2)-\frac{i\pi}{2}\eta(0|A)
\nn\\
&&-\frac{1}{2}\aq\eta'(0|A+im)-\eta'(0|A-im)\cq
+\frac{\pi}{2}\sum_{j=0}^{p}\frac{(-1)^j
m^{2j+1}}{\Ga(j+\fr{3}{2})}A_{p-j}(L)\,.
\label{kkk111-}
\eeq

It should be noted the appearance of other two ``eta-like" non local 
contributions.
This, again, is valid for arbitrary value of the mass $m$.  
Also in the odd dimensional case, there is  agreement with the expressions 
obtained in \cite{dese98-57-7444}, even though, in those papers,
the extra non-local contributions appeared in a different form. 
Eq. (\ref{me1}) leads to
\beq
\eta'(0|A+im)-\eta'(0|A-im)=-2i\int_0^\ii \frac{\sin m t}{t}
\Tr\at \frac{A}{|A|} e^{-t|A|}\ct\,.
\label{517}\eeq
With regard to the large mass limit of the non-local contributions, 
we observe that, from the Voros theorem, one has
\beq
\ze'(0|L+m^2)\simeq\sum_{r=0}^\ii  \Ga(r-p-\frac{1}{2})
A_r(L)m^{2p-2r+1}\,
\eeq
while, from Eqs.~(\ref{boh1}) and (\ref{517}), 
\beq
\lim_{m\to\ii}\aq\eta'(0|A+im)-\eta'(0|A-im)\cq=-i\pi\eta(0|A)
\eeq
easily follows. This means that in the large mass limit
the leading eta contribution cancel and we finally obtain  
\beq
\ze_{+}'(0|K)&\sim&\frac12
\sum_{j=0}^p\aq \Ga(-j-\frac{1}{2})
-\frac{(-1)^j\pi}{\Ga(j+\fr{3}{2})}\cq A_{p-j}(L)m^{2j+1}
+O(1/m)\,.
\eeq

\s{The multiplicative anomaly for the massive Dirac operator}

If four dimensions it is usually assumed that
$\det K=\det K^+=\sqrt{\det K^+K}$ (see, for example, the recent papers
\cite{schu98}). However, a multiplicative anomaly may be present sometimes, 
what will expoil the second equality, 
while the first one could be spoiled by a phase of the determinant. 
This has been recently discussed in detail in \cite{eliz99}, where a 
very simple, specific example in which this
situation occurs has been given (see also the contribution \cite{asada99}). 

To begin with, it should be noted here that, strictly speaking, the Wodzicki 
multiplicative anomaly formula is {\it not} valid for Dirac-like operators, 
since they are (first order) non positive operators. 
However, we still may apply the direct
definition of the  multiplicative anomaly, i.e.    
\beq
a(K^+,K)=\ln \det K^+K-\ln \det K^+-\ln\det K\,.
\label{vr}
\eeq 
The determinant of the differential operator $K^+K=L+m^2$
is defined through the zeta function by
\beq
-\ln\det K^+K=-\ln \det (L+m^2)=\ze'(0|L+m^2)\:,
\eeq
while for $K$ and $K^+$ themselves, we have to set  
\beq
-\ln\det K=\ze_+'(0|K)\:,\hs\hs
-\ln\det K^+=\ze_-'(0|K^+)\:.
\label{kkk3-}
\eeq
The choice $\ze_+$ and $\ze_-$ for $K$ and $K^+$ respectively
is a consequence of the discussion we have carried out at the beginning 
of the previous section.
Now, we have
\begin{description}
\item{\bf Proposition 2.}
For the regularized determinant one gets the nice 
property 
$$\ln \det K^+=(\ln \det K)^{*},$$ 
which is valid for the determinant of non-hermitian matrices.
\end{description}
Coming back to the multiplicative anomaly, we obtain from Eqs. (\ref{kkk3-}),
(\ref{kkk99}) and  (\ref{kkk99-}) a non vanishing result, indeed
\begin{description}
\item{\bf Proposition 3.}
The multiplicative anomaly for the massive Dirac operator
and its adjoint, in a space of even dimension $D=2p$,  reads
\beq
a_{2p}(K^+,K)=2\sum_{j=1}^{p}\frac{(-1)^j m^{2j}c_j}{j!}A_{p-j}(L)\:
\,.\label{coneven}
\eeq
\item{\bf Proposition 4.}
The multiplicative anomaly for the massive Dirac operator
and its adjoint, in a space of odd dimension $D=2p+1$, is given by
\beq
a_{2p+1}(K^+,K)=\pi\sum_{j=0}^{p}(-1)^{j+1}
\frac{m^{2j+1}}{\Ga(j+\fr{3}{2})}\:A_{p-j}(L)\,.
\label{con}
\eeq
\end{description}
$A_{p-j}(L)$ are the Seeley-DeWitt coefficients corresponding
to the operator $L$. 
Of course, when $m\to0$ the multiplicative anomaly vanishes. 
Directly in terms of the determinants of the Dirac operators, we can 
express our results as follows:
\begin{description}
\item{\bf Corollary 1.} In the even dimensional case:
\beq
\left. \frac{\det K}{ \det K^+}\right|_{even}
\ = \ \left( -1 \right)^{\zeta(0|L+m^2)}, \nn 
\eeq
\beq
\left. \frac{\det K}{\sqrt{\det (L+m^2)}}
\right|_{even}
\ = \ \left( -1 \right)^{- \frac{1}{2} \zeta(0|L+m^2)} \, \cdot \,
e^{-\frac{1}{2} a_{2p}(K^+,K)}.
\nn\eeq
\end{description}
The first of the terms in both
expressions coincides with the one obtained by Asada 
in the mathematical literature
\cite{asada99} (after correction of a missprint there). Moreover, we also 
agree with the statement that the remaining anomaly contribution in our
second result could only be given by a
linear combination of Seeley-DeWitt coefficients ---as these terms are 
excluded from the regularized definition of the zeta function determinant
adopted in \cite{asada99}.  

In the odd dimensional case, 
the eta function of the Laplacian undertakes the role of the zeta function, 
In fact, we have
\begin{description}
\item{\bf Corollary 2.}
\beq
\left. \frac{\det K}{ \det K^+}\right|_{odd}
\ = \ \left( -1 \right)^{-\eta(0|A)}\, \cdot \,
e^{- \at \eta'(0|K)-\eta'(0|K^+)\ct}
,\nn 
\eeq
\beq
\left. \frac{\det K}{\sqrt{\det (L+m^2)}}
\right|_{odd} \ = \ \left( -1 \right)^{- \frac{1}{2} \eta(0|A)} \, \cdot \,
e^{ - \frac{1}{2} \at \eta'(0|K)-\eta'(0|K^+)\ct} \, \cdot \,
e^{-\fr{1}{2}a_{2p+1}(K^+,K)}.\nn
\eeq
\end{description}
\s{Some explicit examples}

{\bf Example 2.}
As a simple example of exact computation, let us consider the first order
 operator $A=P-a$, with
$P=-i\partial_\tau$, acting on $S^1$, and $a$ a constant which can be
interpreted as the mean value of a unidimensional periodic vector potential.
We may define the domain of $A$ as  consisting of periodic or antiperiodic
functions with period $2\pi$. Let us consider  periodic (twisted) spinors.
The antiperiodic case is recovered through the replacement
 $a \to a+1/2$. The related eta-function, for large $\Re s$, is
\beq
\eta(s|A)=\sum_{n} \frac{n-a}{|n-a|}(n-a)^{-\fr{s}{2}}\,.
\eeq
The analytic extension near $s=0$ can be obtained
by using the spectral theorem,
\beq
\Tr Ae^{-tA^2}=\sum_{n} (n-a)e^{-t(n-a)^2}\,.
\eeq
The Poisson resummation formula yields
\beq
\Tr Ae^{-tA^2}=-2\left(\frac{\pi}{t}\right)^{3/2} \sum_{n=1}^\ii n
e^{-\fr{\pi^2n^2}{t}}\sin 2\pi na\,.
\eeq
A direct calculation gives
\beq
\eta(s|A)=-2  \sqrt \pi \frac{\Ga(1-s)}{\Ga(\fr{s+1}{2})}
\sum_{n=1}^\ii (\pi n)^{2s-1}\sin 2\pi na\,
\label{eta2}
\eeq
and, as a consequence, the eta-invariant reads, for example for $0 <a<1$, 
\beq
\eta(A)=\eta(0|A)=-2  \frac{1}{\pi}
\sum_{n=1}^\ii  n^{-1}\sin 2\pi na=2a-1\,.
\label{eta3}
\eeq
For  a generic $a$ 
\beq
\eta(A)=\eta(0|A)==2a-2[a]-1\,.
\label{eta4}
\eeq
Note that this is defined modulo $k$, $k$ integer
(i.e., it is a discontinuos and periodic function of
$a$, of period 1). This result is obtained
from classical Fourier analysis
(see, e.g., \cite{knopp}). On the other hand (and this is actually
non-trivial), the result (\ref{eta3}) is the same that one obtains
by analytical continuation  of the eta function $\eta(s|A)$ to $s=0$
---what is clear, e.g. from the corresponding expression to be found 
in \cite{actor}.

Furthermore, let us check the formula (\ref{boh1}). For the sake of 
simplicity let us take $0 <a<1$, but the analysis can be extended to a 
generic $a$. The spectral theorem gives
\beq
\Tr \at \frac{A}{|A|} e^{-t|A|}\ct &=&\sum_{n} \frac{n-a}{|n-a|}e^{-t|n-a|}=
-e^{at}+\at  e^{at}-e^{-at} \ct \sum_{n=1}^\ii e^{-nt} \nn\\
&=& \frac{\sinh (a-1/2)t}{\sinh t/2}\,.
\eeq
Thus
\beq
\lim_{t \to 0}\Tr \at \frac{A}{|A|} e^{-t|A|}\ct =2a-1=\eta(0|A)\,.
\eeq
We  can  also compute 
\beq
\int_0^\ii dt \frac{\sin mt}{t}\Tr \at \frac{A}{|A|} e^{-t|A|}\ct
&=&
\int_0^\ii dt \frac{\sin mt}{t}\frac{\sinh (a-1/2)t}{\sinh t/2}\nn\\
&=&
\arctan{\at \tanh(\pi m) \tan(a-1/2) \ct}\,.
\eeq
In the limit $m \to \ii$  one gets $\pi (a-1/2)$, in complete agreement with
the general result of Sec. 5. Eq. (\ref{me11}) is verified too. 
\medskip

\noindent {\bf Example 3.}
As another example, let us   consider a spinor 
field defined on a Riemann surface of 
genus $g>1$, namely a topologically non trivial 
2-dimensional curved manifold with constant curvature
(see, for example, \cite{zerb93-27-19,byts96-266-1}). 

Being this a 2-dimensional problem, it is sufficient to deal with the 
spinorial 
Laplacian. Recall that a Riemannn surface is locally homeomorphic to the 
2-dimensional hyperbolic plane  $ H^2$. Thus, making use of the half plane 
model for $ H^2$, one has a metric tensor
\begin{equation}
g_{\mu\nu}=y^{-2}\delta_{\mu\nu}.
\label{G}
\end{equation}
The covariant derivative for the 2-component spinor field is given by
\begin{equation}
\nabla_{\mu} \psi=\partial _{\mu}\psi +\omega_{\mu} \psi,
\label{++}
\end{equation}
where the spin connection is determined by the 2-bein $e^a_\mu$, namely 
\begin{equation}
\omega_{\mu}=\frac{1}{8}e_{a\nu}\nabla_\mu e^\nu_b[\Gamma^a,\Gamma^b],
\label{o}
\end{equation}
in which the 2-dimensional Dirac matrices are given in terms of Pauli matrices
\beq
\Gamma^1=\sigma_1, \hs\hs
\Gamma^2=\sigma_2, \hs\hs
\Gamma=\sigma_3.
\label{***}
\eeq
The Dirac operator is $A= i\gamma^\mu\nabla_\mu$, and its square $L$  reads 
\begin{equation}
L=-y^2(\partial^2_x+\partial^2_y)+iy\partial_x\sigma_3-\frac{1}{4}.
\label{ac}
\end{equation} 
We shall deal with 
Weyl spinors, namely the ones which are eigenvectors with eigenvalues
$\pm1$ of $\sigma_3$. Recall that a spinor structure can be introduced in 
the following way
(see for example \cite{hejh76b,dhok86-104-537,sarn87-110-113}).
Let us consider a $\tilde{\Gamma}$,  
containing the element $-1$ and such that $\tilde{\Gamma}/\{\pm\}=\Gamma$. 
Here $\Gamma$ is a stricly hyperbolic discrete 
subgroup of $ PSL(2,R)$.
Define a character $ \chi : \tilde{\Gamma}\rightarrow
 \{ \pm 1\}$, $\chi(-I)=-1$. It is possible to show that there exist $2^{2g}$
inequivalent choices for the spinor characters.
The set of Weyl spinors may be identified as automorphic 
functions of weight $\pm1$. 

The two dimensional self-adjoint operators  given by
\begin{equation}
D_{\pm1}=-y^2(\partial^2_x+\partial^2_y)\pm iy\partial_x
\label{a2}
\end{equation} 
are the related  spinorial Laplacians, while the corresponding squares 
of the Dirac 
operators acting on the automorphic functions of weight $\pm1$ are given by 
$L^{\pm}=D_{\pm1}-1/4$, where the last constant term comes from the constant
curvature, normalized to minus one, of the Riemann surface. From now on, we 
shall  consider only Weyl spinors of weight 1.

As is known, we need the analytic continuation of
the zeta function associated with the differential operator $D_1$.
Such an analytic continuation can be obtained by means of the Selberg trace 
formula. In our case, it reads \cite{hejh76b}
\begin{equation}
\sum_{n=0}^{\infty}h(r_n)=
\frac{V(F)}{2}\int_{-\infty}^\infty h(r)\psi_{2}(r)\ dr
+\sum_{\wp}\sum_{k=1}^{\infty}\frac{\chi(P(\gamma)^{k})}
{S_2(k;l_{\gamma})}\hat h(kl_{\gamma}),
\label{STF}
\end{equation}
where $F$ is a fundamental domain for $\Gamma$, $h(r)$ is an even function,
analytic on a strip and such that  its Fourier transform $\hat h$ is 
exponentially bounded, $(r_n)^2=\lambda_n-1/4$,  
$\lambda_n$ being the eigenvalues of the $D_1$ operator. Furthermore,
$\gamma$ is an element of the conjugacy class associated with the length of 
the closed
geodesic $l_\gamma$, $\wp$ is a set of primitive closed geodesics, and
\beq
\psi_2(r)=\frac{1}{2\pi}r\coth\pi r, \hs\hs
S_2(k;l_{\gamma})&=&\frac{2\sinh (kl_{\gamma}/2)}{l_{\gamma}},\nn\eeq
$\chi(P(\gamma))$ being the $2^{2g}$ characters which define all topological 
inequivalent spinor fields. 

It is convenient to work with the operator $L(\la)=D_1-\la$ and  put
$\de^2=1/4-\la$. Thus, when $\de \to 0$, $L(\la) \to L^+$. 
The  final result for the analytical continuation reads
\begin{eqnarray}
\zeta(s| L(\lambda)&=&\frac{V(F)}{2\pi}\aq \frac{\Gamma(s-1)}
{\Gamma(s)}\delta^{2-2s}
 +2\int_0^{\infty}dr\frac{r(r^2+\delta^2)^{-s}}{e^{2\pi r}-1}\cq\nonumber\\
&&\hs+\frac{\sin(\pi s)}{\pi}\int_0^\infty[t(t+2\delta)]^{-s}
\frac{d}{dt}\ln Z(t+1/2+\delta)\  dt\,,
\label{Z}
\end{eqnarray}
where the Selberg zeta function is given by
\begin{equation}
Z_(s)=\prod_{\{l_{\gamma}\}}
\prod_{k}\aq 1-\chi(l_{\gamma})e^{-(s+k)l_{\gamma}} \cq \,.
\label{Z2}
\end{equation}
Let us consider the massive case. Here $\de^2=m^2$ and, from Eq. (\ref{Z}), we have
\beq
\ze(0|L^++m^2)=\frac{V(F)}{2\pi}\at \frac{\ze_R(2)}{2\pi}-m^2 \ct\,,
\eeq
where $\ze_R(z)$ is the Riemann zeta function and 
\begin{equation}
\ze'(0|L^++m^2)=-C(m)-\ln Z(1/2+m)\,,
\label{ff+}
\end{equation}
with
\begin{equation}
C(\de)=\frac{V(F)}{\pi}\aq \int_0^{\infty}drr\frac{\log (r^2+\delta^2)}
{e^{2\pi r}-1}+\frac{\delta^2}{4}(1-\log \delta^2)\cq\,.
\label{C}
\end{equation}

In the massless case, zero modes may be present. Denoting by  $N$ the number 
of zero modes of the Dirac operator and taking into 
account these zero modes in the evaluation of the zeta function regularized 
determinat of the operator $L^+$, one has the explicit results:
\beq
\ze(0|L^+)=\frac{V(F)}{2\pi} \frac{\ze_R(2)}{2\pi}\,,
\eeq
and
\begin{equation}
\ze'(0| L^+)=- \frac{V(F)}{\pi}\ze_R'(-1)-
\ln \at \frac{Z^{(2N)}(1/2)}{2N!}\ct\,,
\label{zzz}
\end{equation}
which show the power of our procedure.

According to the results of the previous sections, these formulas
allow now for the explicit evaluation of 
the functional determinants of the Dirac operators $A$ and $A+im$. 
Finally we would like to mention that in Refs. \cite{bvz99,bw99} one can find
other explicit examples of computation of the eta invariant related to 
3-dimensional compact hyperbolic manifolds.

\s{Conclusions}

We have obtained in this paper explicit formulas for the 
 zeta and eta--functions associated with the massless and massive Dirac 
operators in an arbitrary manifold without boundary, as a first step 
towards the calculation of the corresponding determinants, what has been
done subsequently. A number of explicit examples have then been considered.  
We have proceeded with the calculation in complete rigor and detail,
since both are necessary when trying to elucidate the long standing issues
concerning the phase of the Dirac operator determinant and the usual 
assumptions (in the physical literature) that  the determinants for plus
and minus $im$ are coinciding and equal on its turn to the square root of the 
massive Laplacian determinant. 

We have here  
settled down these issues (in the negative), once and for all,
working all the time in the zeta-function (and eta-function) domain and using
both powerful mathematical theorems and explicit formulas and calculations 
(including resummations of the mass expansions). Our main results are
given in the form of several propositions that provide  clear answers 
 to the above conjectures (see Sec. 6). 

A remarkable finding is the fact that,
contrary to what had been ordinarily supposed,  the ambiguity present in the 
definition of the associated fermion functional 
determinant in the massless case does actually 
 disappear in the massive case. This
gives rise to a phase of the Dirac determinant that agrees with a very recent 
calculation in the mathematical literature (after an improvement of 
the last).
Our very explicit results for the  multiplicative anomaly of the determinants
of the massive Dirac operators (with $+im$ and $-im$, respectively) are
also in full agreement with recent calculations, in 
the overlapping situations.  All those provide solid  checks of our 
method and of our final formulas.

\s{Appendix}

In this Appendix, we shall consider a covenient algorithm, which is 
equivalent to the calculus of $\Psi$DOs, in order to construct a parametrix 
for the resolvents of Dirac-like operators.
We follow the classical Ref. \cite{atiya} (see also \cite{cogn90}).
For the sake of simplicity, let us consider  $K=A+im$, as a  massive Dirac
operator on a flat (compact) manifold. Let us introduce the
$\ep$-symbol
\beq
\si_\ep(K)=e^{i\fr{kx}{\ep}}Ke^{-i\fr{kx}{\ep}}=K(x,p+\fr{k}{\ep})\,,
= p\!\!/- A\!\!/+\frac{ k\!\!/}{\ep}+im,
\eeq
where  $\ep$ is a
real parameter, defined in order to determine recursively the
parametrix. 
The $\ep$-symbol for the resolvent is given by
\beq
\at \si_\ep(K)-\frac{\la}{\ep}\ct R_\la(\ep;x,k)=I\,.
\label{a1}
\eeq
Here, $I$ is the identity operator and $\la$ a complex number. 
The parametrix for the symbol of the complex power is obtained putting 
$\ep=1$. It is given by
\beq
\si(K^{-s})(x,k)=\frac{1}{2\pi i}\int_{\Ga_\theta} d\la \la^{-s} R_\la(x,k)
\,,
\label{res}
\eeq
where $\Ga_\theta$ is a suitable cut in the $\la$ complex plane. Recall that 
the Fourier transform of (\ref{res}) gives the parametrix for
the kernel of the complex power $K^{-s}$.
One may solve  Eq. (\ref{a1}) with the ansatz
\beq
R_\la(\ep;x,k)=\sum_{l=1}^\ii \ep^l R_l(x,k)\,.
\eeq
As a result, one has the recurrence relations
\beq
(k\!\!/-\la)R_1=I \,, \hs
(k\!\!/-\la)R_l+KR_{l-1}=0\,, \hs l=2,3,...\,.
\eeq
Thus, formally one obtaines
\beq
R_\la(x,k)&=&\frac{1}{k\!\!/-\la}\at I
-K\frac{1}{k\!\!/-\la}
+K\frac{1}{k\!\!/-\la} K \frac{1}{k\!\!/-\la}
\cp\nn\\&&\hs\hs\hs\hs\ap
-K\frac{1}{k\!\!/-\la}K\frac{1}{k\!\!/-\la}
K\frac{1}{k\!\!/-\la}+...\ct \,.
\label{res1}
\eeq
The choice of the cut in the complex $\la$ plane depends  on the 
meromorphic structure of $R_\la(x,k)$. If $K$ is hermitian, 
the poles are on the real
axis and  two possible inequivalent cuts can be chosen in the lower or 
upper complex plane, as discussed in the Introduction. 
If a mass is present or, in general,
 the operator $K$ is non-hermitian, the situation changes. 
As an illustration, we may consider the free case, namely
$K^{(0)}=p\!\!/+im$. A 
formal resummation gives
\beq
R_\la^{(0)}(x,k)
=\frac{1}{k\!\!/+im-\la}=\frac{k\!\!/-im+\la}{k^2-(im-\la)^2}\,.
\label{res0}
\eeq
Here the poles are located in the upper complex plane and one is forced to 
choose the lower cut in  Eq. (\ref{res}).  This argument is also valid in
the interacting case, because the ``free" contribution can be additively
separated in Eq. (\ref{res1}).
 
On the other hand, in the case of the adjoint operator $K^+$, the only
change is $im \to -im$ and it is plain that, in this case, the cut must 
be chosen in the upper
complex $\la$ plane, as argued in Sec. 4.

\ack{
EE is indebted with the members of the 
Department of Mathematics and of the Center for 
Theoretical Physics
 at MIT, specially with Dan Freedman and
Robert Jaffe, for very kind hospitality during the time this work
was carried out. We also thank Andrei Bytsenko for interesting discussions.
This work has been supported by the cooperative agreement
INFN (Italy)--DGICYT (Spain).
EE has been financed also by CIRIT (Generalitat de Catalunya), grant
1998BEAI400208, and by DGICYT (Spain), project PB96-0925.
}

\end{document}